\documentclass[aps,prl,twocolumn,preprintnumbers,superscriptaddress,amsmath]{revtex4}

\usepackage{amssymb}
\usepackage{txfonts}
\usepackage{amssymb}
\usepackage{graphicx}
\usepackage{subfigure}
\usepackage{dcolumn}
\usepackage{bm}
\usepackage{color}
\usepackage{upgreek}
\usepackage{bbm}
\usepackage{lineno}

\begin{document}
%\linenumbers
%\title{Arbitrary multi-qubit quantum entanglement generation with high-dimensional encoding}
\title{On-chip generation of multi-qubit graph states with high-dimensional encoded single photons}
%A scheme for resource-efficient multi-qubit entanglement state generation with single photons path-encoded
%A scheme for resource-efficient photonic multi-qubit entanglement state generation
%Integrated and resource-efficient multi-qubit entanglement state generation with single photons

\author{Lan-Tian Feng}
\author{Bo-Hao Zhang}
%\author{Bo-Yu Xu}
%\author{Xin-Yu Song}
\author{Di Liu}
\affiliation
{Laboratory of Quantum Information, University of Science and Technology of China, Hefei 230026, China.}
\affiliation{CAS Synergetic Innovation Center of Quantum Information $\&$ Quantum Physics, University of Science and Technology of China, Hefei 230026, China.}
\affiliation{Anhui Province Key Laboratory of Quantum Network, University of Science and Technology of China, Hefei 230026, China.}
%\author{Yu-De Ge}
\author{Pan Gong}
\author{Yu-Yang Ding}
\affiliation{Hefei Guizhen Chip Technologies Co., Ltd., Hefei 230000, China.}
\author{Guo-Ping Guo}
%\author{Wei Chen}
\author{Guang-Can Guo}
\author{Xi-Feng Ren\footnote[2]{renxf@ustc.edu.cn}}
\affiliation
{Laboratory of Quantum Information, University of Science and Technology of China, Hefei 230026, China.}
\affiliation{CAS Synergetic Innovation Center of Quantum Information $\&$ Quantum Physics, University of Science and Technology of China, Hefei 230026, China.}
\affiliation{Anhui Province Key Laboratory of Quantum Network, University of Science and Technology of China, Hefei 230026, China.}
%\affiliation{Hefei National Laboratory, University of Science and Technology of China, Hefei 230088, China.}

\begin{abstract}
Photonic multi-qubit entanglement is key to optical quantum information processing, particularly universal quantum computing. Yet multi-photon sources suffer from low emission efficiency, making single-photon high-dimensional encoding an appealing alternative.
%Multi-photon entanglement is a scarce resource.
%for optical quantum information processing, particularly for universal quantum computing.
% Consequently, higher dimensions and various degrees of freedom of photons are introduced to encode multiple qubits.
%However, the conversion between different degrees of freedom is complex, and the power of high-dimensional encoding is poorly explored.
%However, multi-photon sources typically exhibit low emission probabilities.
%Consequently, the utilization of single photons to encode more information is attractive.
%
Here we propose an explicit and resource-efficient high-dimensional encoding approach to achieve the target multi-qubit quantum state. 
%We show that various multi-qubit entangled states can be constructed using single photons. , named layer-by-layer qubit encoding,
% and resource-efficient
The technically challenging preparation of multi-photon quantum states is replaced by single-photon operations involving high-dimensional expansion, routing, and multi-layered quantum measurement.
%straightforward
Besides, each photon in the resource multi-photon quantum state can be used to encode multiple qubits in a distributed manner, and a larger entangled state will be constructed. 
We demonstrate this approach using programmable photonic integrated circuits, where multi-qubit graph states—including the Greenberger-Horne-Zeilinger state and the cluster state—are generated and characterized. 
We additionally demonstrate the Grover search algorithm using the single-photon cluster state.
Our findings unlock a novel route towards diverse entangled state generation with photons and advance large-scale and universal photonic quantum information processing.
\end{abstract}
\maketitle

%\section*{1. INTRODUCTION}

%\noindent {\bf Quantum state tomography.} 

%光量子信息处理有着多种优势，例如低环境要求，芯片集成，光纤互联网络等。
Photonics has developed into one of the crucial physical platforms for quantum information processing, owing to its low environmental demands, ease of chip integration, and widespread availability of fiber optic networks.
On this platform, the weak photon–photon interactions make multi-qubit gates challenging and motivate the development of various multi-qubit entangled states as core resources for photonic quantum technologies.
To achieve valuable quantum applications, a substantial number of entangled qubits are required, each of which should be independently addressable.
However, multi-photon sources typically exhibit low emission probabilities, rendering multi-photon entanglement a scarce resource.
Although heralded detection and feedback multiplexing techniques \cite{Bartolucci2023,Aghaee2025} have been proposed to mitigate this challenge, they will result in a significant increase in overhead and therefore remain a long-term goal.
An alternative approach, more feasible in the near term, is to utilize single photons to encode as much information as possible.

Photons possess numerous degrees of freedom (DOFs) for encoding quantum information, and many of these degrees are extendable to high dimensions.
These features allow us to encode multiple qubits using single photons, which proves to be more resistant to decoherence and experimentally more resource-efficient than their multi-photon counterparts \cite{Azzini2020}.
%the singe-particle entanglement do satisfy the usual theoretical criteria for entanglement {Phys. Rev. A, 54, 3824}
%In this scenario, the quantum correlation is shared within the same photon, leading to the observed intra-particle entanglement.
%Importantly, this single-photon entanglement satisfies the usual theoretical criteria for entanglement .
%
In this scenario, the quantum correlation within a single photon gives rise to observed intra-particle entanglement, satisfying conventional entanglement criteria \cite{Bennett1996}.
Furthermore, when combined with multi-photon sources, single-photon entanglement has become a key resource widely utilized in quantum information science.
For example, encoding quantum information across multiple DOFs has enabled the realization of the hyper-entangled 10-qubit Schrödinger cat state \cite{Gao2010} and 18-qubit Greenberger-Horne-Zeilinger (GHZ) state \cite{Wangxl2018}, as well as the demonstration of quantum algorithms including gate teleportation \cite{Huang2004}, one-way quantum computation \cite{Chen2007,Tame2014}, error-correcting codes \cite{Bell2014} and 3-qubit logic gates \cite{Kagalwala2017}.
This multi-DOF conversion technology has also been explored in quantum photonic integrated circuits \cite{Feng2016}, and more recently, a chip-scale polarization-spatial-momentum quantum SWAP gate has been demonstrated \cite{Cheng2023}.
%the number of photon  is finite, an
Despite these advances, coherent conversion between different DOFs often introduces significant losses and degrades fidelity—presenting a major obstacle to the large-scale integration of this technology into optical chips.

\begin{figure*}[t]
\centering
\includegraphics[width=14.0cm]{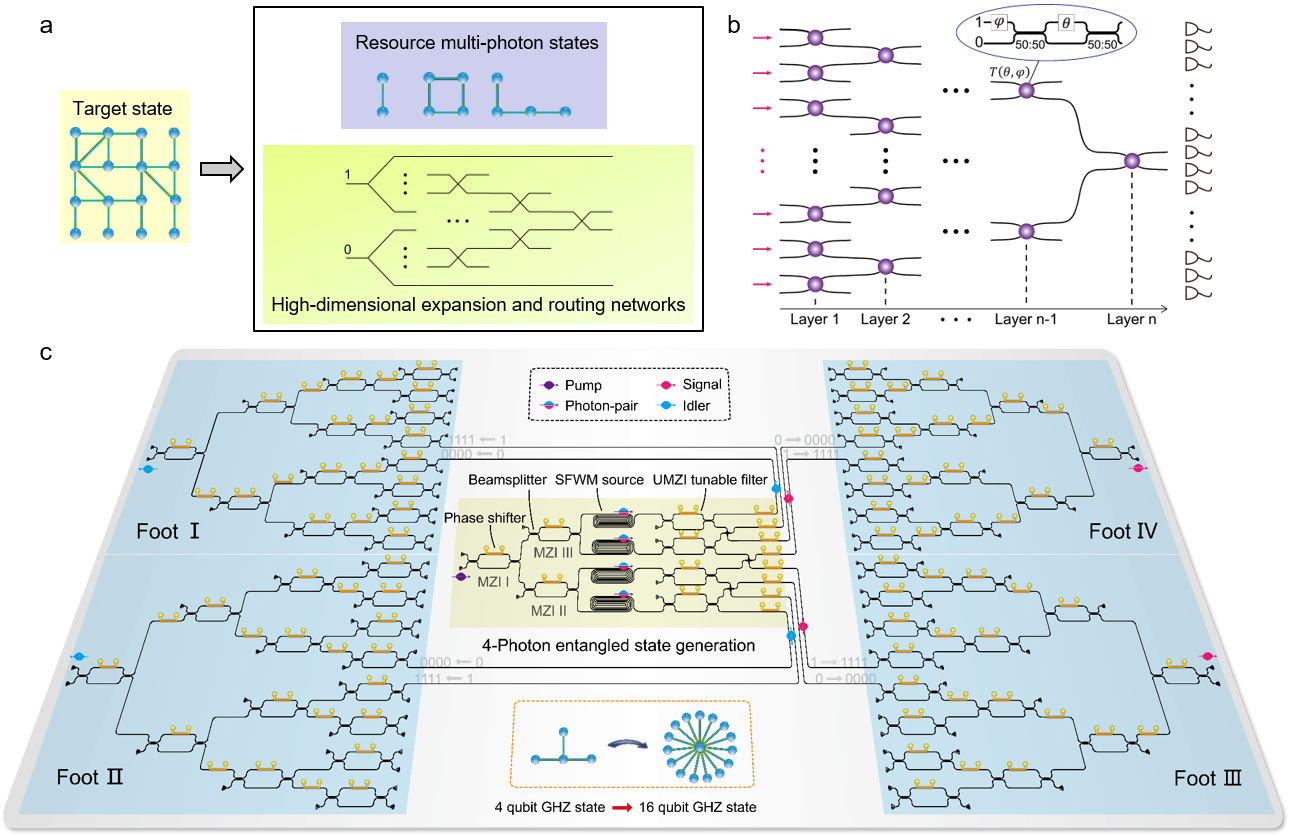}
\caption {\textbf{Conceptual illustration of the proposed approach and circuit.} \textbf{a,} To prepare the target state, we can select appropriate resource multi-photon states and deploy customized high-dimensional expansion and routing networks. \textbf{b,} The layer-by-layer qubit measurement scheme. For $n$-qubit quantum state preparation with single photons, a $n$-layer measurement structure is arranged, and each layer can achieve complete measurement of the corresponding qubit. \textbf{c,} The schematic diagram of photonic integrated circuit for 4-photon 16-qubit GHZ state preparation. Owing to the consistency in the forms of GHZ states with different qubit numbers, high-dimensional expansion and routing operations are unnecessary, and the path dimensions can be directly relabeled.} 
\label{fig1}
\end{figure*}
%MZI: Mach-Zehnder interferometer; UMZI: unbalanced Mach-Zehnder interferometer.
%As we can see, although encoding multiple qubits using single photons has achieved various entangled states preparation and quantum information applications, the further increase in the number of qubits still faces great difficulties. 
%To further increase the scale of the system, an effective framework to guide the preparation of arbitrary multi-qubit quantum states while minimizing resource consumption is becoming essential.
%More importantly, an effective framework to guide the preparation of arbitrary multi-qubit quantum states is needed.

Beyond utilizing multiple DOFs of single photons, high-dimensional encoding based on a single DOF, such as paths, transverse spatial modes, or time-frequency bins, is gaining popularity for encoding more information \cite{Babazadeh2017,Hu2018,Cozzolino2019,Erhard2020,Chi2023}. 
This approach alleviates the functional requirements for manipulating devices across distinct DOFs, thereby facilitating large-scale integration \cite{Wang2018,Hu2020,Bao2023,Yu2025}. 
In recent years, high-dimensional techniques have enabled the demonstration of cluster state generation with more than 9 qubits \cite{Lib2024}, error-protected on-chip qubits \cite{Vigliar2021}, and chip-to-chip quantum gate teleportation \cite{Feng2024}.
%However, high-dimensional encoding with single photons generally requires operations in the Hilbert space up to $2^n$ dimensions ($n$ is the qubit number), and further single- and multi-qubit gates and measurements, which introduces significant resource challenges.  
However, high-dimensional encoding with single photons requires operations in a Hilbert space of up to $2^n$ dimensions ($n$ being the number of qubits), imposing considerable resource constraints.
Consequently, a method to reduce resource consumption in multi-qubit state preparation is becoming increasingly essential.
%while how to minimize resource consumption in multi-qubit entanglement preparation remains largely unexplored.

%
Here, we propose an effective approach for multi-qubit quantum state preparation based on high-dimensional encoding.
%By combining quantum state preparation and state measurement, 
%the multi-qubit state construction process is converted to a multilayer quantum measurement process. 
Leveraging this approach, diverse multi-qubit states can be generated from single-photon inputs, and multi-photon sources can be harnessed to construct larger-scale entangled states.
%According to this method
As examples, we have developed two photonic integrated circuits to respectively prepare the 4-photon 16-qubit GHZ state and the single-photon 4-qubit cluster state.
The genuine entanglement is witnessed for the 10-qubit GHZ state and verified by quantum state tomography for the 4-qubit cluster state.
%Furthermore, one-way quantum computation with the single-photon 4-qubit cluster state is demonstrated.
%
These results pave the way for multi-qubit photonic quantum information applications.

Our approach is illustrated in Fig. 1a. To prepare the target state, high-dimensional expansion and routing operations are performed on photons of the resource multi-photon state, which are determined by state form differences between two states.
Though the target state is generated via these operations, the critical challenge lies in how to implement single-qubit measurements to verify state quality or enable subsequent quantum operations.
For this aim, a layer-by-layer qubit measurement scheme is proposed (Fig. 1b), and measurement of each bit corresponds to the operation of the corresponding layer.
%The schematic diagram of constructing large-scale entangled state by combining multi-photon entangled source and single-photon multi-qubit encoding techniques is shown in Fig. 1a.
%After the multi-photon entanglement source is obtained, each photon in the source is extended by the measurement device to encode more qubits.
%To encode $n$ qubit with single photons, a $2^{n}$ dimensional Hilbert space and subsequent single-qubit and multi-qubit gates are generally required, which will lead to a very complex circuit. 
%To overcome this difficulty, we have proposed an explicit and resource-efficient multilayer measurement framework for multi-qubit state preparation, and the architecture is shown in Fig. 1b. 
%In this method, the $n$-layer measurement structure is arranged for any $n$-qubit pure quantum state preparation. 
%
In the scheme, each layer contains $2^{(n-m)}$ ($m=1,...,n$ denotes the layer number) measurement nodes, and each node can achieve any single-qubit measurement, 
\begin{equation}
T(\theta,\varphi)=e^{i\frac{\theta}{2}}\begin{bmatrix}
\cos\frac{\theta}{2} & -ie^{i\frac{\varphi}{2}}\sin\frac{\theta}{2}\\
-i\sin\frac{\theta}{2}& e^{i\frac{\varphi}{2}}\cos\frac{\theta}{2}
\end{bmatrix}.
\end{equation} 
%(inset in Fig. 1b)
Here, qubit number in the target state is ordered from right to left.
That is, if we have a state of $n$ qubits and label each qubit $1\to n$, qubit $n$ is the leftmost one and qubit $1$ is the rightmost one. 
Thus, the number of layers in Fig. 1b corresponds one-to-one with the number of qubits, and two dimensions of each node encode 0 and 1 states of the corresponding qubit.
By configuring nodes in the same layer $m$ with the same operation $T(\theta_{m},\varphi_{m})$, we will achieve a complete measurement of the corresponding qubit in the target state.
%, and will be set as the same measurement.  

We have validated the high-dimensional encoding approach with integrated silicon photonic circuits, which have achieved a very large scale \cite{Bao2023,Harris2016} and high visibility \cite{Psi2025} with dual-rail encoding and have become a scalable platform for quantum information processing \cite{silverstone2016silicon,wang2020integrated,Feng2022}. 
The conceptual layout of the integrated silicon photonic circuit for 16-qubit GHZ state preparation is shown in Fig. 1c. 
We adopt path of photons as the high-dimensional DOF, and the method can be readily extended to other high-dimensional encoding systems.
With four spiral single-mode silicon waveguides as photon-pair sources, we first prepare a 4-photon GHZ state. 
Then each photon is used to encode four qubits, and a 4-photon 16-qubit GHZ state is achieved. 

In the experiment, a 200 GHz bandwidth pulsed laser centered at 1550.12 nm acts as the pump. 
It is coupled into the chip using a grating coupler and split coherently into four paths using three Mach-Zehnder interferometers (MZIs I-III) configured to act as 50:50 beam splitters.
On each path, a 1.3 cm long single-mode spiral silicon waveguide is pumped to generate signal-idler photon pairs with the spontaneously four-wave mixing process \cite{feng2019}. 
After passing through the sources, signal and idler photons in four paths are separated using unbalanced MZIs (UMZIs). 
Subsequently, a waveguide crossing network is used to swap photon's paths, and a 4-photon GHZ state is generated \cite{Adcock2019,Llewellyn2020,Lee2024}.
Upon inputting each photon of the entangled state into the four-layer measurement module (Foots I–IV), the entire system is expanded to form a 16-qubit GHZ state.
%By programming phase shifters in the measurement structure, universal quantum projection measurement can be achieved for any qubit in this state.
Configuring phases in each layer enables arbitrary single-qubit measurement capabilities for the corresponding qubit.
The photons are coupled out from the chip via grating couplers, and then they are delivered into four single-photon detectors for detection and further time correlation analysis. 
Based on the nature of the UMZI filter \cite{Feng2024}, we select signal and idler photons with respective central wavelengths of 1538.19 nm and 15662.23 nm and 100 GHz bandwidth. 
More details about the optical circuit and experimental setup are given in the Supplementary Information, Sec. I.

\begin{figure}[t]
\centering
\includegraphics[width=8.0cm]{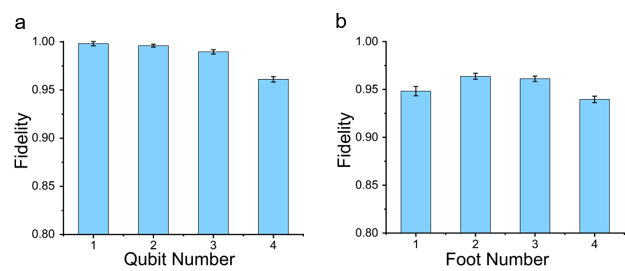}
\caption {\textbf{Results of the single-photon multi-qubit GHZ state.} \textbf{a,} The fidelity of single-photon encoded GHZ state with different qubit numbers. \textbf{b,} The fidelity of single-photon encoded 4-qubit GHZ state with different measurement modules (Foots I-IV). The errors are obtained from 300 simulation calculations with the experimental data subjected to Gaussian statics.}
\label{fig2}
\end{figure}

To verify the effectiveness of our approach, we first use herald single-photons to verify the reliability of a single measurement module (Foot III). 
The herald single photons are generated by one packaged silicon nanowire photon-pair source. 
%(see Supplementary Information for more information)
They are coupled into the chip, split coherently into two paths
using MZI I and routed to Foot III. 
%Although there are only single photons, b
By setting phases in each layer, we can obtain the GHZ state $|G_m\rangle=\frac{1}{\sqrt{2}}(|0\rangle^{\otimes m}+|1\rangle^{\otimes m})$ with the number of qubits up to 4. 
For example, by guiding photons directly through the first two layers, and the last two layers are used to analyze the state, we will obtain the 2-qubit GHZ state.
That is, the expected qubit number determines how many layers are used to perform measurement operations.
%
%By configuring phases in each layer, we can achieve arbitrary single-qubit measurement capabilities, which allows us to perform all projection measurements $\{\left|0\right\rangle,\left|1\right\rangle,\left|0\right\rangle+\left|1\right\rangle,\left|0\right\rangle+i\left|1\right\rangle\}^{\otimes m}$ that required to reconstruct the density matrix $\hat\rho$ of the state.
%
We perform the quantum state tomography test for single-photon multipartite GHZ states. 
With the fidelity between the ideal density matrix $\hat\rho_{\rm{ideal}}$ of the state and the measured density matrix $\hat\rho_{\rm{mea}}$ defined as $F=\rm{Tr}(\hat\rho_{\rm{mea}}\hat\rho_{\rm{ideal}})$, where $\rm{Tr}$ represents the trace, the fidelities for the GHZ state with different qubit numbers are shown in Fig. 2a.   
Due to the high quality of on-chip single-qubit operations, the fidelity of the 4-qubit GHZ state has reached 0.961(0.003), which is currently the highest fidelity of the 4-qubit states achieved in integrated optical chip systems.
Similarly, we perform the 4-qubit quantum state tomography test for other measurement modules (Foots I, II, and IV), and the results are shown in Fig. 2b.
The average fidelity of 0.953(0.004) for the 4-qubit GHZ state of four modules is achieved.

\begin{figure}[t]
\centering
\includegraphics[width=8cm]{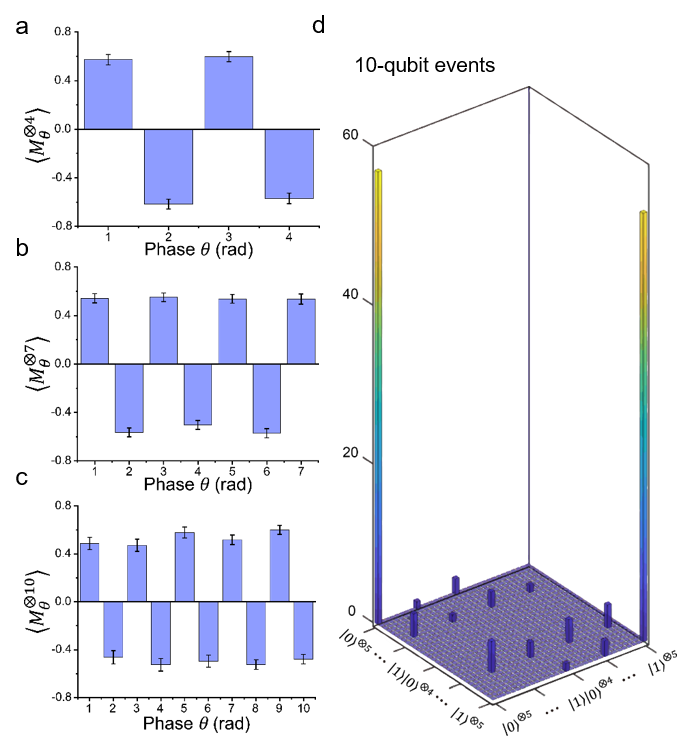}
\caption {\textbf{Results of the four-photon multi-qubit GHZ state.} \textbf{a-c,} The expectation values of $M_k$ for the 4-qubit (a), 7-qubit (b) and 10-qubit (c) GHZ state. %They show the measured coherence of the GHZ states. 
The coherence values of the 4-qubit, 7-qubit, and 10-qubit GHZ states are calculated to be 0.589(0.020), 0.543(0.015), and 0.516(0.014), respectively. \textbf{d,} Fourfold coincidence counts measured in the 0/1 basis for the 10-qubit GHZ state. The populations of the
$|0\rangle^{\otimes 10}$ and $|1\rangle^{\otimes 10}$ terms are calculated to be 0.815(0.024).}
\label{fig3}
\end{figure}

\begin{figure*}[t]
\centering
\includegraphics[width=15.0cm]{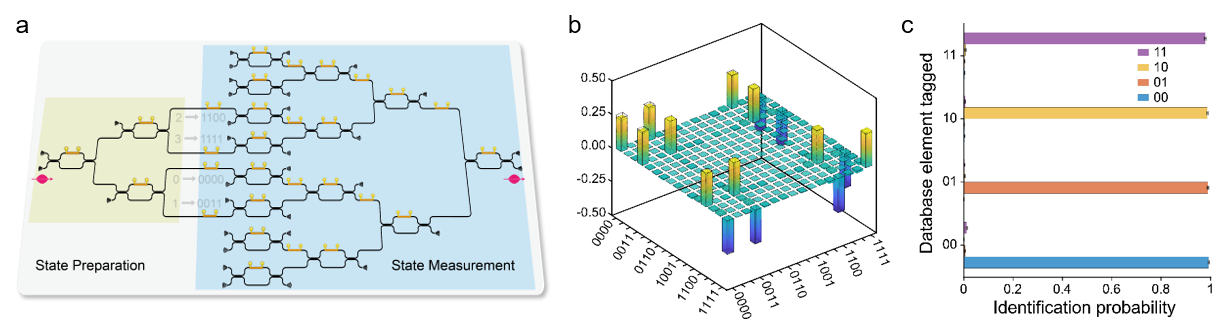}
\caption {\textbf{Results of the single-photon 4-qubit cluster state.} \textbf{a,} The schematic diagram of photonic integrated circuit for single-photon 4-qubit cluster state preparation. The four path dimensions are relabeled, each corresponding to a product term of the cluster state. \textbf{b,} Real part of the measured density matrix of the cluster state. Crystal clear bars represent the density matrix for the ideal cluster state. The imaginary part of the measured density matrix is negligible. \textbf{c,} The identification probability for different labeled elements.}
\label{fig4}
\end{figure*}

Next, we characterize the multipartite entanglement with more qubits.
%with a repetition rate of 100 MHz
A pulsed laser is coupled into the chip and used to pump the spiral silicon waveguides coherently to prepare quantum photonic sources \cite{zhang2019generation}. 
When every measurement module (Foot I-IV) is used only to encode a single qubit, that is, only the last layer of each module is manipulated, we obtain the 4-qubit GHZ state with four photons.
%, which has been achieved in some recent work \cite{Adcock2019,Llewellyn2020,Lee2024}The advantage in this work is that
Since each photon can be used to encode up to four qubits by programming the circuit, we can obtain up to 16-qubit entanglement.   
%thus 
To verify the genuine multipartite entanglement, we use the entanglement witness approach \cite{Gühne2007}, and the operator for the multipartite GHZ state is
\begin{equation}
W_{G_m}=\frac{I}{2}-|G_m\rangle\langle{G_m}|=\frac{I}{2}-\frac{A}{2}-\frac{1}{m}\sum_{k=0}^{m}(-1)^{k}M_k,
\end{equation} 
where $m$ is the qubit number,  $A=|0\rangle^{\otimes{m}}\langle{0}|^{\otimes{m}}+|1\rangle^{\otimes{m}}\langle{1}|^{\otimes{m}}$ and $M_k=[\cos(\frac{k\pi}{m})\sigma_x+\sin(\frac{k\pi}{m})\sigma_y]^{\otimes{m}}$, ($k=1,...,m$) are local measurement operators.
For the $n$-qubit state, each local measurement operator is constructed from the results obtained from $2^{n}$ different measurement settings, which are achieved by programming on-chip phases. 

We have conducted entanglement witness tests on 4-qubit, 7-qubit, and 10-qubit GHZ states, respectively.
Considering that there are multiple combinations that can achieve 7-qubit and 10-qubit GHZ states, here, we use Foot III only to encode four qubits to obtian the 7-qubit GHZ state and Foots III and IV to respectively encode four qubits to obtain the 10-qubit GHZ state.
The measurement results are shown in Fig. 3. 
With a total of $n+1$ measurement operators, we obtain $\langle W_{G_{m=4,7,10}}\rangle$ values of -0.225(0.015), -0.189(0.011) and -0.166(0.014), respectively.
With high statistical significance ($>11\sigma$), the result confirms the
genuine 10-qubit entanglement, which is the largest entangled state demonstrated so far in photonic chips.  
If only classical dual-rail encoding is exploited, the 10-photon GHZ state would exhibit a count rate of $1.1\times10^{-15}$ Hz, a value so low that state verification becomes fundamentally infeasible.
%our high-dimensional 10-qubit GHZ state is $\sim$14 orders of magnitude more efficient.   
%Although the 16-qubit GHZ state is not verified, 
%
In contrast, the brightness of our 10-qubit GHZ state reaches 0.154 Hz.
For the verification of entanglement with more qubits, the reduction of measurement time becomes extremely necessary (see Supplementary Information, Sec. III).
One adopted method is increasing the number of photonic detectors and thus simultaneously performing 4-photon event measurements on multiple measurement settings \cite{Wangxl2018} . 
%

%More importantly, . To show that our approach is suitable for preparing any classes of pure and entangled multipartite states, 
%
Finally, we employ our approach to generate cluster states with single photons, which serve as the core resources for measurement-based quantum computing \cite{Raussendorf2001}. 
The 4-qubit cluster state we prepared has the form of
\begin{equation}
|C_4\rangle=\frac{1}{2}(|0000\rangle+|0011\rangle+|1100\rangle-|1111\rangle).
\end{equation} 
This state has been demonstrated with four photons \cite{Walther2005,Adcock2019,Prevedel2007} and two photons entangled in both polarization and spatial mode \cite{Chen2007,Vallone2008}. 
Here, we demonstrate it with single photons. 
Figure 4a shows the schematic diagram of the circuit to prepare the single-photon 4-qubit cluster state.
The heralded single photons are entered into the chip and divided equally into four parts, and each part corresponds to one term of the state $|C_4\rangle$ (marked in Fig. 4a).
Following quantum state preparation, the state is fed into the four-layer measurement module for state analysis.
%Compared to the measurement module in Fig. 1a, the structure is simplified, and only the necessary interferometers which are involved in the modulation of the photon state in each layer are included. 
%which is not detrimental to the measurement results.
%

We perform quantum state tomography to measure the state density matrix of the 4-qubit cluster state, and a fidelity of 0.991(0.004) is obtained, which confirms the high quality.
The Bell inequality of the cluster state is further verified with the following operator \cite{Scarani2005}:
\begin{equation}
S=\sigma_z\mathbbm{1}\sigma_x\sigma_x+\sigma_x\sigma_y\sigma_y\sigma_x+\sigma_x\sigma_y\sigma_x\sigma_y-\sigma_z\mathbbm{1}\sigma_y\sigma_y.
\end{equation} 
For the cluster state $|C_4\rangle$, the maximal expectation value of $S$ is achieved ($S=4$), in contrast to the bound of $S=2$ for local hidden variable models.
Although single-particle entanglement has no way to achieve the space-like separation like multi-particle entangled states, it could still violate the inequality.
The $S$ parameter is experimentally estimated as 3.921(0.005).
%, which evidently violates the inequality.
%
We then demonstrate the Grover search algorithm, which aims to find the marked element in an unsorted database \cite{Grover1997}.
%One-way realization of the Grover search algorithm needs the box cluster state, which distinguishes the cluster $|C_4\rangle$ from an H transformation on every qubit and a swap between qubits 2 and 3. For the cluster $|C_4\rangle$, we need to measure the qubits along the 0/1 basis.
In one-way realization, the algorithm is transformed to single-qubit measurements on the cluster state.
With the 4-qubit cluster state, we are able to find the marked element with a single quantum search in the database with four entries $|00\rangle$, $|01\rangle$, $|10\rangle$, and $|11\rangle$ \cite{Chen2007,Walther2005,Prevedel2007}. 
In the experiment, we tag the labeled element on qubits 2,3 and make the readout on qubits 1,4.
With the measurement settings of qubits 2,3 consistent with the labeled elements, we have measured all the combinations of qubits 1,4 along the 0/1 basis.
The experimental results are sketched in Fig. 4c.
For four different targets, the average identification probability of 0.987(0.003) is achieved.

The above results have already demonstrated the effectiveness of the proposed approach. 
Compared with the commonly adopted gate-based models for preparing multi-qubit entangled quantum states using high-dimensional single photons, our approach achieves resource savings by eliminating complex single-qubit and multi-qubit gate operations (see Supplementary Information, Sec. IV).
%To further increase the qubit number, assessment of minimum resource consumption to prepare different classes of quantum states with single photons is necessary. Here, the minimum means that we care about only the necessary measurement nodes.
%
For a given target state, computer programs can be leveraged to further identify the optimal configurations for the required resource states, as well as high-dimensional network expansion and routing architectures.
Besides, not all measurement nodes within the layer-by-layer qubit measurement scheme are essential; those with no input photons can be eliminated to further simplify the chip structure.
The minimum resource consumption of the measurement structure can be readily associated with the minimal number of product terms $r$ of the target state, which gives an account on the degree of entanglement of a multipartite system \cite{Eisert2001}. 
%Schmidt measure of the target state, . , as demonstrated in Fig. 4a. 
Specifically, the $n$-qubit pure separable state and GHZ state only need $\mathcal{O}(n)$ nodes, the W state needs $\mathcal{O}(n^2)$ nodes, while the cluster state needs $\mathcal{O}(2^n)$ nodes.
Considering the large-scale scalability of integrated photonic chip technology \cite{Bao2023}, it is feasible to achieve single-photon entangled states of hundreds of qubits.
%Furthermore, when combined with the on-chip multi-photon sources, resource consumption will be further reduced.
%
Beyond qubits, the single-qubit measurement node can be extended to high-dimensional qudit measurement, enabling the realization of diverse multipartite high-dimensional entangled states. 
Currently, most of these states remain at the theoretical level primarily because of the challenges in generating such states with multiple particles. 

In conclusion, we have provided a high-dimensional encoding approach for generating multi-qubit states using single photons, which encompasses single-photon operations involving high-dimensional expansion, routing, and multi-layered quantum measurement.
%The layer-by-layer encoding architecture enables facile large-scale scaling while ensuring high fidelity.
%
As an initial proof-of-concept, we developed two distinct photonic integrated circuits, which are designed to respectively prepare the 4-photon 16-qubit GHZ state and the single-photon 4-qubit cluster state.
Both the genuine entanglement of the 10-qubit GHZ state and the high fidelity of the single-photon 4-qubit cluster state validate the effectiveness of our proposed method.
This method is resource-efficient and convenient for large-scale integration, and it enables the preparation of photonic quantum states with more qubits and diverse entangled structures, laying the groundwork for developing dedicated photonic quantum processors.
Therefore, our work provides a novel route towards diverse entangled state generation with photons and advances the realization of large-scale and practical photonic quantum information processing.

\textit{Acknowledgments}—This work was supported by National Key Research and Development Program of China (2022YFA1204704), the National Natural Science Foundation of China (62275240, 62435009, U25D9003, T2325022, U23A2074), the CAS Project for Young Scientists in Basic Research (No. YSBR-049), and the Fundamental Research Funds for the Central Universities. This work was partially carried out at the USTC Center for Micro and Nanoscale Research and Fabrication.

\textit{Data availability}—The data supporting this study are not publicly available but may be obtained from the corresponding author upon reasonable request.

%\section*{REFERENCES}
%\bibliography{ref}

\begin{thebibliography}{99}
\bibitem{Bartolucci2023} S. Bartolucci, P. Birchall, H. Bombín, H. Cable, C. Dawson, M. Gimeno-Segovia, E. Johnston, K. Kieling, N. Nickerson, M. Pant, F. Pastawski, T. Rudolph, and C. Sparrow, Fusion-based quantum computation. Nat. Commun. \textbf{14,} 912 (2023).

\bibitem{Aghaee2025} H. Aghaee Rad, T. Ainsworth, R. N. Alexander, et al, Scaling and networking a modular photonic quantum computer. Nature \textbf{638,} 912--919 (2025).

\bibitem{Azzini2020} S. Azzini, S. Mazzucchi, V. Moretti, D. Pastorello, and L. Pavesi, Single-particle entanglement. Adv. Quantum Technol. \textbf{3,} 2000014 (2020).

\bibitem{Bennett1996} C. H. Bennett, D. P. DiVincenzo, J. A. Smolin, and W. K. Wootters, Mixed-state entanglement and quantum error correction. Phys. Rev. A \textbf{54,} 3824 (1996).

\bibitem{Gao2010} W. B. Gao, C. Y. Lu, X. C. Yao, P. Xu, O. Gühne, A. Goebel, Y. A. Chen, C. Z. Peng, Z. B. Chen, and J. W. Pan, Experimental demonstration of a hyper-entangled ten-qubit Schrödinger cat state. Nature Phys. \textbf{6,} 331--335 (2010).

\bibitem{Wangxl2018} X. L. Wang, Y. H. Luo, H. L. Huang, M. C. Chen, Z. E. Su, C. Liu, C. Chen, W. Li, Y. Q. Fang, X. Jiang, J. Zhang, L. Li, N. L. Liu, C. Y. Lu, and J. W. Pan, 18-qubit entanglement with six photons' three degrees of freedom. Phys. Rev. Lett. \textbf{120,} 260502 (2018).

\bibitem{Huang2004} Y. F. Huang, X. F. Ren, Y. S. Zhang, L. M. Duan, and G. C. Guo, Experimental teleportation of a quantum controlled-NOT gate. Phys. Rev. Lett. \textbf{93,} 240501 (2004).

\bibitem{Chen2007} K. Chen, C. M. Li, Q. Zhang, Y. A. Chen, A. Goebel, S. Chen, A. Mair, and J. W. Pan, Experimental realization of one-way quantum computing with two-photon four-qubit cluster states. Phys. Rev. Lett. \textbf{99,} 120503 (2007).

\bibitem{Tame2014} M. S. Tame, B. A. Bell, C. Di Franco, W. J. Wadsworth, and J. G. Rarity, Experimental realization of a one-way quantum computer algorithm solving Simon’s problem. Phys. Rev. Lett. \textbf{113,} 200501 (2014).

\bibitem{Bell2014} B. A. Bell, D. A. Herrera-Mart\'{i}, M. S. Tame, D. Markham, W. J. Wadsworth, and J. G. Rarity, Experimental demonstration of a graph state quantum error-correction code. Nat Commun. \textbf{5,} 3658 (2014).

\bibitem{Kagalwala2017} K. H. Kagalwala, G. D. Giuseppe, A. F. Abouraddy, and B. E. A. Saleh, Single-photon three-qubit quantum logic using spatial light modulators. Nat. Commun. \textbf{8,} 739 (2017).

\bibitem{Feng2016} L. T. Feng, M. Zhang, Z. Y. Zhou, M. Li, X. Xiong, L. Yu, B. S. Shi, G. P. Guo, D. X. Dai, X. F. Ren, and G. C. Guo, On-chip coherent conversion of photonic quantum entanglement between different degrees of freedom. Nat. Commun. \textbf{7,} 11985 (2016).

\bibitem{Cheng2023} X. Cheng, K. C. Chang, Z. Xie, M. C. Sarihan, Y. S. Lee, Y. Li, X. Xu, A. K. Vinod, S. Kocaman, M. Yu, P. G. Q. Lo, D. L. Kwong, J. H. Shapiro, F. N. C. Wong, and C. W. Wong, A chip-scale polarization-spatial-momentum quantum SWAP gate in silicon nanophotonics. Nat. Photon. \textbf{17,} 656--665 (2023).

\bibitem{Babazadeh2017} A. Babazadeh, M. Erhard, F. Wang, M, Malik, R. Nouroozi, M. Krenn, and A. Zeilinger, High-dimensional single-photon quantum gates: concepts and experiments. Phys. Rev. Lett. \textbf{119,} 180510 (2017).

\bibitem{Hu2018} X. M. Hu, Y. Guo, B. H. Liu, Y. F. Huang, C. F. Li, and G. C. Guo, Beating the channel capacity limit for superdense coding with entangled ququarts. Sci. Adv. \textbf{4,} eaat9304 (2018).

\bibitem{Cozzolino2019} D. Cozzolino, B. Da Lio, D. Bacco, and L. K. Oxenløwe, High-dimensional quantum communication: benefits, progress, and future challenges. Adv. Quantum Technol. \textbf{2,} 1900038 (2019).

\bibitem{Erhard2020} M. Erhard, M. Krenn, and A. Zeilinger, Advances in high-dimensional quantum entanglement. Nat. Rev. Phys. \textbf{2,} 365--381 (2020).

\bibitem{Chi2023} Y. Chi, Y. Yu, Q. Gong, and J. Wang, High-dimensional quantum information processing on programmable integrated photonic chips. Sci. China Inf. Sci. \textbf{66,} 180501 (2023).

\bibitem{Wang2018} J. Wang, S. Paesani, Y. Ding, R. Santagati, P. Skrzypczyk, A. Salavrakos, J. Tura, R. Augusiak, L. Mančinska, D. Bacco, D. Bonneau, J. W. Silverstone, Q. Gong, A. Acín, K. Rottwitt, L. K. Oxenløwe, J. L. O'Brien, A. Laing, and M. G. Thompson, Multidimensional quantum entanglement with large-scale integrated optics, Science \textbf{360,} 285--291 (2018).

\bibitem{Hu2020} X. M. Hu, W. B. Xing, B. H. Liu, Y. F. Huang, C. F. Li, G. C. Guo, P. Erker and M. Huber, Efficient generation of high-dimensional entanglement through multipath down-conversion. Phys. Rev. Lett. \textbf{125,} 090503 (2020).

\bibitem{Bao2023} J. Bao, Z. Fu, T. Pramanik, et al, Very-large-scale integrated quantum graph photonics, Nat. Photon. \textbf{17,} 573--581 (2023).

\bibitem{Yu2025} H. Yu, S. Sciara, M. Chemnitz, N. Montaut, B. Crockett, B. Fischer, R. Helsten, B. Wetzel, T. A. Goebel, R. G. Krämer, B. E. Little, S. T. Chu, S. Nolte, Z. Wang, J. Azaña, W. J. Munro, D. J. Moss, and R. Morandotti, Quantum key distribution implemented with d-level time-bin entangled photons. Nat. Commun. \textbf{16,} 171 (2025).

%\bibitem{Reimer2019} C. Reimer, S. Sciara, P. Roztocki, M. Islam, L. R. Cortés, Y. Zhang, B. Fischer, S. Loranger, R. Kashyap, A. Cino, S. T. Chu, B. E. Little, D. J. Moss, L. Caspani, W. J. Munro, J. Azaña, M. Kues, and R. Morandotti, High-dimensional one-way quantum processing implemented on $d$-level cluster states, Nature Phys. \textbf{15,} 148--153 (2019).

\bibitem{Lib2024} O. Lib, Y. Bromberg, Resource-efficient photonic quantum computation with high-dimensional cluster states. Nat. Photon. \textbf{18,} 1218--1224 (2024).

\bibitem{Vigliar2021} C. Vigliar, S. Paesani, Y. Ding, J. C. Adcock, J. Wang, S. Morley-Short, D. Bacco, L. K. Oxenløwe, M. G. Thompson, J. G. Rarity, and A. Laing, Error-protected qubits in a silicon photonic chip, Nat. Phys. \textbf{17,} 1137--1143 (2021).

%\bibitem{Cao2023} S. Cao, B. Wu, F. Chen,  et al, Generation of genuine entanglement up to 51 superconducting qubits. Nature \textbf{619,} 738--742 (2023).

%\bibitem{Panda2025} D. K. Panda, and C. Benjamin, Designing three-way-entangled and nonlocal two-way-entangled single-particle states via alternate quantum walks. Phys. Rev. A \textbf{111,} 012420 (2025).

\bibitem{Feng2024} L. T. Feng, M. Zhang, D. Liu, Y. J. Cheng, X. Y. Song, Y. Y. Ding, D. X. Dai, G. P. Guo, G. C. Guo and X. F. Ren, Chip-to-chip quantum photonic controlled-NOT gate teleportation. Phys. Rev. Lett. \textbf{135,} 020802 (2025).

\bibitem{Harris2016} N. C. Harris, D. Bunandar, M. Pant, G. R. Steinbrecher, J. Mower, M. Prabhu, T. Baehr-Jones, M. Hochberg, and D. Englund, Large-scale quantum photonic circuits in silicon. Nanophotonics \textbf{5,} 456--468 (2016).

\bibitem{Psi2025} PsiQuantum team. A manufacturable platform for photonic quantum computing. Nature \textbf{641,} 876--883 (2025).

\bibitem{silverstone2016silicon} J. W. Silverstone, D. Bonneau, J. L. O'Brien, and M. G. Thompson, Silicon quantum photonics. \emph{IEEE J. Sel. Top. Quant.} \textbf{22,} 390-402 (2016).

\bibitem{wang2020integrated} J. Wang, F. Sciarrino, A. Laing, and M. G. Thompson, Integrated photonic quantum technologies. Nature Photon. \textbf{14,} 273-284 (2020).

\bibitem{Feng2022} L. Feng, M. Zhang, J. Wang, X. Zhou, X. Qiang, G. Guo, and X. Ren, Silicon photonic devices for scalable quantum information applications. Photon. Res. \textbf{10,} A135--A153 (2022).

\bibitem{feng2019} L. T. Feng, M. Zhang, X. Xiong, Y. Chen, H. Wu, M. Li, G. P. Guo, G. C. Guo, D. X. Dai, and X. F. Ren, On-chip transverse-mode entangled photon pair source. npj Quantum Inf. \textbf{5,} 2 (2019).

%\bibitem{Silverstone2015} J. W. Silverstone, R. Santagati, D. Bonneau, M. J. Strain, M. Sorel, J. L. O’Brien, and M. G. Thompson, Qubit entanglement between ring-resonator photon-pair sources on a silicon chip. Nat. Commun. \textbf{6,} 7948 (2015).

\bibitem{Adcock2019} J. C. Adcock, C. Vigliar, R. Santagati, J. W. Silverstone, and M. G. Thompson, Programmable four-photon graph states on a silicon chip. Nat. Commun. \textbf{10,} 3528 (2019).

\bibitem{Llewellyn2020} D. Llewellyn, Y. Ding, I. I. Faruque, S. Paesani, D. Bacco, R. Santagati, Y.-J. Qian, Y. Li, Y.-F. Xiao, M. Huber, M. Malik, G. F. Sinclair, X. Zhou, K. Rottwitt, J. L. O’Brien, J. G. Rarity, Q. Gong, L. K. Oxenlowe, J. Wang, and M. G. Thompson, Chip-to-chip quantum teleportation and multi-photon entanglement in silicon. Nat. Phys. \textbf{16,} 148--153 (2020).

\bibitem{Lee2024} J. M. Lee, J. Park, J. Bang, Y. I. Sohn, A. Baldazzi, M. Sanna, S. Azzini, and L. Pavesi, Quantum states generation and manipulation in a programmable silicon-photonic four-qubit system with high-fidelity and purity. APL Photon. \textbf{9,} 076110 (2024).

\bibitem{zhang2019generation} M. Zhang, L. T. Feng, Z. Y. Zhou, Y. Chen, H. Wu, M. Li, S. M. Gao, G. P. Guo, G. C. Guo, D. X. Dai, and X. F. Ren, Generation of multiphoton quantum states on silicon. Light Sci. Appl. \textbf{8,} 41 (2019).

\bibitem{Gühne2007} O. Gühne, C.-Y. Lu, W.-B. Gao, and J.-W. Pan, Toolbox for entanglement detection and fidelity estimation. Phys. Rev. A \textbf{76,} 030305 (2007).

\bibitem{Raussendorf2001} R. Raussendorf and H. J. Briegel, A one-way quantum computer. Phys. Rev. Lett. \textbf{86,} 5188--5191 (2001).

\bibitem{Walther2005} P. Walther, K. J. Resch, T. Rudolph, E. Schenck, H. Weinfurter, V. Vedral, M. Aspelmeyer, and A. Zeilinger, Experimental one-way quantum computing. Nature \textbf{434,} 169--176 (2005).

\bibitem{Prevedel2007} R. Prevedel, P. Walther, F. Tiefenbacher, P. Böhi, R. Kaltenbaek, T. Jennewein, and A. Zeilinger, High-speed linear optics quantum computing using active feed-forward. Nature \textbf{445,} 65--69 (2007).

\bibitem{Vallone2008} G. Vallone, E. Pomarico, F. D. Martini, and P. Mataloni, Active one-way quantum computation with two-photon four-qubit cluster states. Phys. Rev. Lett. \textbf{100,} 160502 (2008).

%\bibitem{Zhang2016} C. Zhang, Y. F. Huang, C. J. Zhang, J. Wang, B. H. Liu, C. F. Li, and G. C. Guo, Generation and applications of an ultrahigh-fidelity four-photon Greenberger-Horne-Zeilinger state. Opt. Express \textbf{24,} 27059--27069 (2016).

\bibitem{Scarani2005} V. Scarani, A. Acín, E. Schenck, and M. Aspelmeyer, Nonlocality of cluster states of qubits, Phys. Rev. A \textbf{71,} 042325 (2005).

\bibitem{Grover1997} L. K. Grover, Quantum mechanics helps in searching for a needle in a haystack. Phys. Rev. Lett. \textbf{79,} 325--328 (1997).

%\bibitem{Michler1999} M. Michler, H. Weinfurter, and M. Żukowski, Experiments towards falsification of noncontextual hidden variable theories. Phys. Rev. Lett. \textbf{84,} 5457--5461 (1999).

\bibitem{Eisert2001} J. Eisert and H. J. Briegel, Schmidt measure as a tool for quantifying multiparticle entanglement. Phys. Rev. A \textbf{64,} 022306 (2001).

%\bibitem{Erhard2018} M. Erhard, M. Malik, M. Krenn, and A. Zeilinger, Experimental Greenberger–Horne–Zeilinger entanglement beyond qubits. Nature Photon. \textbf{12,} 759--764 (2018).

%\bibitem{Hu2025} X. M. Hu, C. X. Huang, N. d'Alessandro, G. Cobucci, C. Zhang, Y. Guo, Y. F. Huang, C. F. Li, G. C. Guo, X. Gao, M. Huber, A. Tavakoli, and B. H. Liu, Observation of genuine high-dimensional multi-partite non-locality in entangled photon states. Nat. Commun. \textbf{16,} 5017 (2025).

\end{thebibliography}

%Conflict of interest: The authors declare no competing interests. 
%Data Availability: The data supporting this study are not publicly available but may be obtained from the corresponding author upon reasonable request.

%Author contributions—XFR supervised and managed the project. LTF, BHZ and DL built the experimental setup with assistance of GPG and GCG. LTF designed the device, calibrated the system, carried out the experiments, and analyzed the experimental data. PG and YYD developed the chips. All authors contributed to the writing of the manuscript.

\end{document}